\documentclass[twocolumn,twoside,slac_two]{revtex4}
\usepackage{graphicx}
\usepackage{fancyhdr}
\begin{document}

\title{Parsec-Scale Radio Properties of Gamma-Ray Emitting Blazars}

%

\author{J. D. Linford, G. B. Taylor}
\affiliation{Department of Physics and Astronomy, University of New Mexico, MSC07 4220, Albuquerque, NM 87131-0001, USA}
\author{R. W. Romani}
\affiliation{Department of Physics, Stanford University, Stanford, CA 94305, USA}
\author{J. F. Helmboldt}
\affiliation{Naval Research Laboratory, Code 7213, 4555 Overlook Ave. SW, Washington, DC 20375, USA}
\author{A. C. S. Readhead, R. Reeves, J. L. Richards}
\affiliation{Astronomy Department, California Institute of Technology, Mail Code 247-17, 1200 East California Boulevard, Pasadena, CA 91125, USA}

\begin{abstract}
\small
The parsec-scale radio properties of blazars detected by the Large Area Telescope (LAT) on board the \textit{Fermi Gamma-ray Space Telescope} have been investigated using observations with the Very Long Baseline Array (VLBA).  Comparisons between LAT and non-LAT detected samples were made using contemporaneous data.  In total, 232 sources were used in the LAT-detected sample.  This very large, radio flux-limited sample of active galactic nuclei (AGN) provides insights into the mechanism that produces strong $\gamma$-ray emission.  It has been found that LAT-detected BL Lac objects are very similar to the non-LAT BL Lac objects in most properties, although LAT BL Lac objects may have longer jets.  The LAT flat spectrum radio quasars (FSRQs) are significantly different from non-LAT FSRQs and are likely extreme members of the FSRQ population.  Contemporaneous observations showed a strong correlation, whereas no correlation is found using archival radio data.  Most of the differences between the LAT and non-LAT populations are related to the cores of the sources, indicating that the $\gamma$-ray emission may originate near the base of the jets (i.e., within a few pc of the central engine).  There is some indication that LAT-detected sources may have larger jet opening angles than the non-LAT sources.  Strong core polarization is significantly more common among the LAT sources, suggesting that $\gamma$-ray emission is related to strong, uniform magnetic fields at the base of the jets of the blazars.   Observations of sources in two epochs indicate that core fractional polarization was higher when the objects were detected by the LAT.  Included in our sample are several non-blazar AGN such as 3C84, M82, and NGC~6251.  
\end{abstract}
\normalsize
\maketitle

\thispagestyle{fancy}


\section{INTRODUCTION}
The Large Area Telescope (LAT; \cite{LAT}) on board the \textit{Fermi Gamma-ray Space Telescope} is a wide-field telescope covering the energy range from about 20 MeV to more than 300 GeV.  It has been scanning the entire $\gamma$-ray sky approximately once every three hours since July of 2008.  The $\gamma$-ray bright sources from the \textit{Fermi} LAT First Source Catalog (1FGL; \cite{1FGL}) that have been identified with radio sources are most often associated with blazars (685 of 1451 sources).  These blazars typically are strong, compact radio sources which exhibit flat radio spectra, rapid variability, compact cores with one-sided parsec-scale jets, and superluminal motion in the jets \cite{mar06}.  The dominance of blazars continues in the \textit{Fermi} LAT Second Source Catalog (2FGL; \cite{2FGL}), compiled using the first two years of LAT data.

Several very long baseline interferometry (VLBI) programs, such as the Monitoring Of Jets in AGN with VLBA Experiments (MOJAVE; \cite{MOJAVE}) observing at 15 GHz, the Boston University program observing at 22 and 43 GHz \cite{mar10}, and the Tracking AGN with Austral Milliarcsecond Interferometry (TANAMI; \cite{TANAMI}) observing at 8.4 and 22 GHz, along with the LAT collaboration themselves \cite{abdo11BLLac} are in place to monitor the radio jets from the brightest blazars such as 3C273, BL Lac, etc.  Our sample has a flux limit roughly an order of magnitude below the MOJAVE survey and so allows us to probe the extensions of the radio core/$\gamma$-ray properties down to a fainter population.

\section{SAMPLE DEFINITION}
Our contemporaneous radio sample is made up of Very Long Baseline Array (VLBA) observations made in 2009 and 2010, all at a frequency of 5 GHz.  We made new observations of 90 sources from the VLBA Imaging and Polarimetry Survey (VIPS; \cite{VIPS}) and new 5 GHz observations of 7 sources in the MOJAVE sample.  We refer to this VIPS follow-up sample as ``VIPS+''.  We also observed 135 non-VIPS, non-MOJAVE  sources selected from the Fermi 1FGL as sources which were associated with a source in the Combined Radio All-Sky Targeted Eight GHz Survey (CRATES; \cite{CRATES}) with high ($\geq$80\%) probability and had a flux density of at least 30 mJy in CRATES.  We refer to this sample of 135 new objects as ``VIPS++''.  The VIPS++ sample expands the number of LAT sources in our sample and includes objects outside of the original VIPS target RA and DEC, which were limited to the region covered by the Sloan Digital Sky Survey Data Release 5 (see \cite{VIPS} more information).  Of our 232 sources, 95 are BL Lac objects, 107 are FSRQs, and 30 are other types of AGN (radio galaxies, AGN of unknown type, and 1 starburst galaxy).  Any object that is not a BL Lac object or FSRQ we classify as `AGN/Other'.  The optical classifications are taken from the LAT First Catalog of AGN (1LAC; \cite{1LAC}).  See \cite{Lin12} for more information on this contemporaneous sample.

To build a non-LAT detected sample for comparison, we excluded all LAT sources from VIPS, leaving 1018 objects in our non-LAT sample.  Of these 1018 objects, 24 are BL Lacs, 479 are FSRQs, and 515 are AGN/Other types (radio galaxies or AGN of uncertain type).  The optical types for VIPS sources were adopted from the Candidate Gamma-Ray Blazar Survey (CGRaBS; \cite{CGRaBS}).  It should be noted that the non-LAT sample is not contemporaneous with LAT observations.

\section{GAMMA-RAY AND RADIO FLUX}
The LAT $\gamma$-ray fluxes are reported in several energy bands in both 1FGL and 2FGL.  To create total $\gamma$-ray fluxes, we combined the fluxes from 3 bands: 100-300 MeV, 300 MeV - 1 GeV, and 1-100 GeV.  The fluxes were added and uncertainties were added in quadrature.  However, some sources had only upper limits to their fluxes in certain bands.  If a source's reported fluxes in one or two bands were upper limits, we use 1/2 the reported flux as the uncertainty in that band as the upper limits are given as 2-sigma results \cite{1FGL}.

\subsection{GAMMA-RAY - RADIO CORRELATION}

\begin{figure}[t]
\centering
\includegraphics[width=70mm]{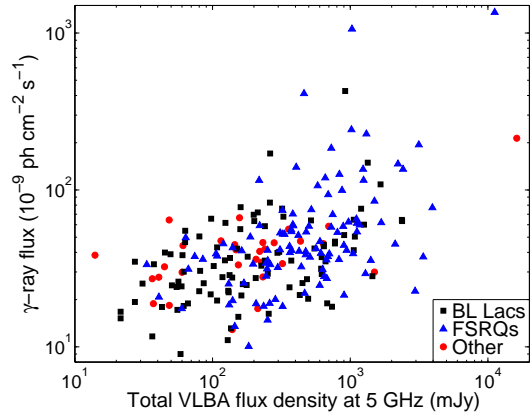}
\caption{LAT $\gamma$-ray flux (100 MeV - 100 GeV) vs. total VLBA radio flux density at 5 GHz.  The $\gamma$-ray fluxes are from 1FGL and are in units of $10^{-9}$ photons cm$^{-2}$ s$^{-1}$.  The black squares are BL Lacs, the blue triangles are FSRQs, and the red circles are radio galaxies and unclassified objects.} \label{FF_nerr}
\end{figure}

In Figure~\ref{FF_nerr} we plot the 1FGL LAT flux versus the total VLBA flux density at 5 GHz.  Again, the LAT fluxes are broadband fluxes from 100 MeV to 100 GeV.  

We used the nonparametric Spearman test \cite{press} to look for correlation between the LAT flux and the total VLBA flux density.  The Spearman $\rho_{s}$ values we found were 0.467 for the BL Lacs and 0.510 for the FSRQs.  The probabilities for getting the same $\rho_{s}$ values from random sampling were 2$\times$10$^{-6}$ for the BL Lac objects and 2$\times$10$^{-8}$ for the FSRQs.  Therefore, the LAT flux correlates very strongly with the total VLBA flux density.  The BL Lac objects do not correlate quite as strongly, especially considering that we found a marginal correlation between total VLBA flux density and redshift for these objects.  Still, this provides solid evidence that objects with higher radio flux density also produce more $\gamma$-ray flux.

In \cite{Lin11}, we found that there was no strong correlation between radio flux density and $\gamma$-ray flux.  However, our flux densities for those observations were prior to the launch of \textit{Fermi}.  Furthermore, it is likely that the radio flux density increases during episodes of $\gamma$-ray flaring \cite{kov09}.  Our newer, larger, and more contemporaneous set of observations are more appropriate for making these comparisons.

\subsection{LAT VERSUS NON-LAT RADIO FLUX DENSITIES}

\begin{figure}[b]
\centering
\includegraphics[width=70mm]{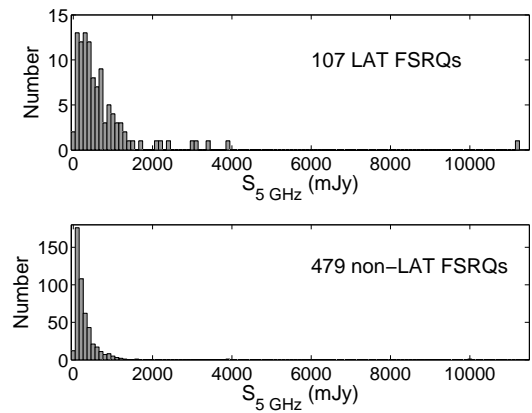}
\caption{Distributions of LAT (top) and non-LAT (bottom) FSRQ total VLBA flux density at 5 GHz.} \label{fsrq_s5}
\end{figure}

The median total VLBA flux density at 5 GHz was 177 mJy for the LAT BL Lacs and 221 mJy for the non-LAT BL Lacs.  The LAT FSRQ appeared to have significantly higher flux densities than the non-LAT FSRQs, with a median of 467 mJy for LAT FSRQs compared to a median of only 191 mJy for non-LAT FSRQs.  The LAT AGN/Other objects also had higher 5 GHz flux densities than their non-LAT sources counterparts.  
The Kolmogorov-Smirnov (K-S) test \cite{press} probability that the LAT and non-LAT BL Lac objects belong to the same parent population is about 38\%, although we should note that we have a very small sample of non-LAT BL Lac objects.  The K-S test probability that the LAT and non-LAT FSRQs are related is only 3$\times$10$^{-12}$.  See Figure ~\ref{fsrq_s5} for a plot of the FSRQ total VLBA 5 GHz flux density distributions.

\subsection{GAMMA-RAY LOUDNESS}

For a measure of ``$\gamma$-ray loudness,''we use the ratio of the $\gamma$-ray luminosity to the radio luminosity as described in \cite{Lis11}.  However, unlike \cite{Lis11}, we do not have multiple observations of our sources from which to determine a median radio luminosity.  Instead, we only use a single VLBA observation to calculate a luminosity.  Due to the variable nature of these blazars, this may not be the best representation of the sources' actual average luminosity.  For the $\gamma$-ray luminosity, we used the average fluxes reported in the 1FGL and the 2FGL.  Whenever possible, we used the redshifts reported in 1LAC.  For sources with no redshift in 1LAC, we searched the NASA/IPAC Extragalactic Database (NED).

All of our sources were significantly $\gamma$-ray loud, with $\gamma$-ray/radio luminosity ratios of at least 400 using both 1FGL and 2FGL data.  It was reported in \cite{Lis11} that the MOJAVE group saw a significant difference between the $\gamma$-ray loudness distributions for BL Lac and FSRQs.  We did not find any indication of this in our data.  The K-S test returned a 8\% probability that the 1FGL BL Lac objects and FSRQs were drawn from the same parent population.  For the 2FGL data, the K-S test result was 18\%.

\section{PARSEC-SCALE RADIO PROPERTIES}

\subsection{CORE BRIGHTNESS TEMPERATURE}

\begin{figure}[t]
\centering
\includegraphics[width=70mm]{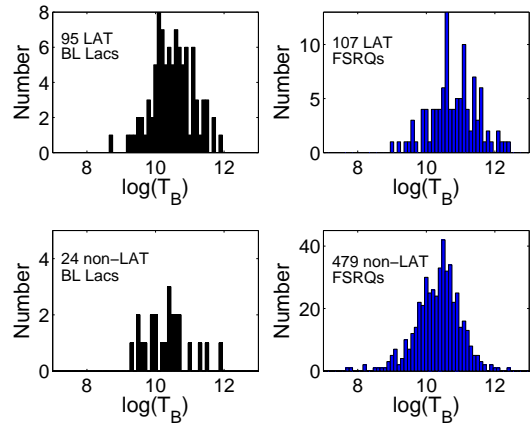}
\caption{Distributions of LAT (top) and non-LAT (bottom) core brightness temperatures.  The BL Lac object distributions are in black on the left.  The FSRQ distributions are in blue on the right.} \label{ctb_dis}
\end{figure}

We looked at the distributions of the core brightness temperatures for our LAT and non-LAT objects.  The LAT FSRQs were significantly different from their non-LAT counterparts.  The K-S test gave a $6\times10^{-8}$ chance that the two distributions were drawn from the same parent population.  The median core brightness temperatures were $6.4\times10^{10}$ K for the LAT FSRQs and $2.5\times10^{10}$ K for the non-LAT FSRQs, indicating that the LAT FSRQs are more extreme sources.  The BL Lac objects did not show a significant difference between the LAT and non-LAT populations.  See Figure ~\ref{ctb_dis} for the distributions of core brightness temperatures for the BL Lac objects and FSRQs.  A difference between the core brightness temperature distributions was also reported in \cite{kov09}, although they did not separate objects by optical type.

\begin{figure}[b]
\centering
\includegraphics[width=70mm]{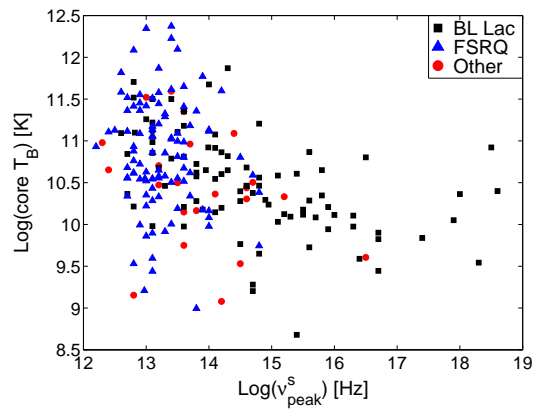}
\caption{Core brightness temperature versus peak synchrotron frequency for LAT-detected sources.}\label{ct_nup}
\end{figure}

We also looked for correlation between core brightness temperature and the frequency at which the synchrotron radiation is at a maximum ($\nu^{S}_{peak}$) for our LAT-detected objects.  Whenever possible, we used published values of $\nu^{S}_{peak}$ (from \cite{Aa11}, \cite{AbdoSED}, \cite{Niep06}, and \cite{Niep08}).  For those sources without published $\nu^{S}_{peak}$, we used the estimation method described in \cite{AbdoSED}, based on the average radio-optical and optical-x-ray spectral indices published in \cite{2LAC}.  If we could not find a published value or use the estimation method, we obtained data from NED and fit a parabola in log space.  We only found a significant correlation for the BL Lac objects.  The Spearman test resulted in a $\rho_{S}$ of $-0.4$ and a $p$ of $10^{-4}$.  See Figure ~\ref{ct_nup} for a plot of core brightness temperature versus peak synchrotron frequency.  It looks very similar to the results presented in \cite{Lis11}.

\subsection{CORE FRACTIONAL POLARIZATION}

\begin{figure}[t]
\centering
\includegraphics[width=70mm]{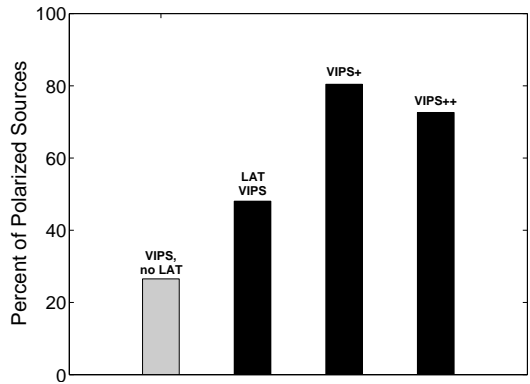}
\caption{Percent of sources with detected core fractional polarization in various observing campaigns.  VIPS observations were made during and prior to 2006.  VIPS+ observations were made in 2009-2010, VIPS++ observations were made in 2010.  Gray indicates the sample with no LAT-detected sources, black indicates LAT samples.}\label{polbar}
\end{figure}

In order for a source to be considered ``polarized'' we required that it have a polarized flux of at least 0.3\% of the Stokes $I$ peak value (to avoid leakage contamination) and have at least a 5$\sigma$ detection (compared to the noise in polarized flux density).  See \cite{Lin12} for more discussion on the polarization poperties of our sources.  We found that the LAT sources are polarized much more often than the non-LAT sources (see Figure ~\ref{polbar}).  The abundance of polarized sources among the $\gamma$-ray emitting population indicates that strong, uniform magnetic fields at the base of the jets are somehow related to the production of $\gamma$-rays.  Also, 48 of the 75 polarized sources which were in both VIPS and VIPS+ showed an increase in core fractional polarization during LAT detection.

\subsection{JET OPENING ANGLE}

We measured the a mean apparent opening half angle for each source with core-jet morphology following the procedure described in \cite{Taylor07}.  This method relies on averaging the apparent opening angles found for each jet component in a source.  That is, the apparent opening angle is measured for several (more than 2) jet components.  The final mean opening angle for the source is the average of all of the opening angles for the individual jet components.

\begin{figure}[t]
\centering
\includegraphics[width=70mm]{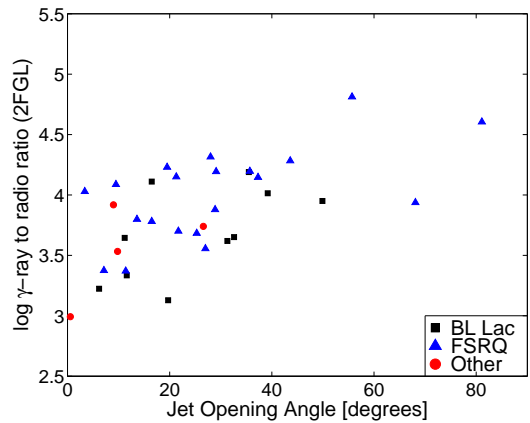}
\caption{2FGL $\gamma$-ray to radio luminosity ratio versus apparent jet opening angle.  The 1FGL data did not show a correlation.}\label{G_oa2}
\end{figure}

The MOJAVE group reported in \cite{Lis11} that they found a non-linear correlation between the apparent opening angle and the $\gamma$-ray loudness.  We also found a correlation, however only when we used the 2FGL data.  When we performed the Spearman test on all 34 sources with both a measured opening angle and a 2FGL $\gamma$-ray flux, we found a $\rho_{S}$ of 0.57 and a $p$ of 0.00044.  When we break the sources up by type, we only see a tentative correlation for the FSRQs ($\rho_{S}$=0.58, $p$=0.0088), however the sample sizes are very small.  We only had 36 objects with opening angle and $\gamma$-ray loudness ($G_{r}$) measurements in 1FGL, and 34 in 2FGL.  The MOJAVE group's sample in \cite{Lis11} is well over 100 sources.  See Figure~\ref{G_oa2} for a plot of the 2FGL $\gamma$-ray loudness versus opening angle.  We should also note that the MOJAVE group in \cite{Lis11} made their measurements of the apparent opening angle using mean sizes of jet components over several epochs, not the mean of multiple apparent opening angles from several components in a single observation as we did.  The fact that MOJAVE is a monitoring program makes it better suited to investigating the apparent (and intrinsic) opening angles than our sample.

\subsection{JET CHARACTERISTICS}

\begin{figure}[t]
\centering
\includegraphics[width=70mm]{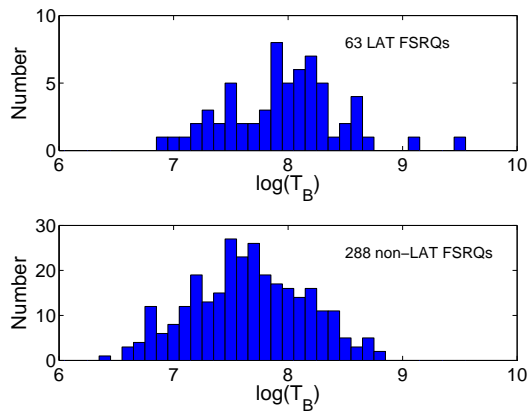}
\caption{Distributions of jet brightness temperatures for LAT (top) and non-LAT (bottom) FSRQs.}\label{fsrq_jt}
\end{figure}

We measured the change in jet position angle (``jet bending'', see \cite{Helm08}) and the length of the jet (the maximum separation between the core component and the jet components).  We applied the K-S test to each of these measurements to see if the LAT and non-LAT distributions were different.  We did not find any significant differences for either of these properties.  However, we do note that the LAT BL Lac objects have a ``long-jet'' morphology (a jet with an angular size of at least 6 milliarseconds) more often than the non-LAT BL Lac objects (see \cite{Lin12}).  Our measurements of the jet brightness temperature (the brightness temperature of the brightest jet component), on the other hand, did produce an interesting result.  We found that the LAT FSRQs had significantly higher jet brightness temperatures than the non-LAT FSRQs.  The K-S test returned a $10^{-5}$ chance that the LAT and non-LAT distributions were drawn form the same parent population.  The FSRQ jet brightness temperature distributions are shown in Figure \ref{fsrq_jt}.

\section{CONCLUSIONS}

Our LAT BL Lac object sample is nearly 4 times the size of our non-LAT BL Lac object sample.  We only find 3 minor differences between the two: (1) the LAT BL Lac objects are polarized more often, (2) the LAT BL Lacs are more often of a ``long-jet'' morphology, and (3) the LAT BL Lac objects \textit{may} have larger apparent jet opening angles.  This leads us to suspect that all BL Lac objects are emitting $\gamma$-rays.

Our LAT and non-LAT FSRQ samples are often very different.  The LAT FSRQs appear to be extreme sources.  They tend to have high radio flux density, high core and jet brightness temperatures, and (maybe) large apparent jet opening angles.  The LAT FSRQs are also polarized much more often than the non-LAT FSRQs.

The correlations found between the radio flux density and $\gamma$-ray flux indicate that the synchrotron and inverse Compton radiation are related.  Therefore, the $\gamma$-rays should originate in the jets, where we know the synchrotron-emitting electrons are located.  However, because most of the differences between the LAT and non-LAT populations are associated with the cores of our blazars, we suspect the $\gamma$-rays are emitted at the base of the jets (i.e., within a few parsecs of the central engine).  The prevalence of polarized cores among the LAT-detected sample indicates that strong, uniform magnetic fields play a role in the production of the $\gamma$-rays.

\bigskip 
\begin{acknowledgments}
The National Radio Astronomy Observatory is a facility of the National Science Foundation operated under cooperative agreement by Associated Universities, Inc.  
This work made use of the Swinburne University of Technology software correlator, developed as part of the Australian Major National Research Facilities Programme and operated under license.
The NASA/IPAC Extragalactic Database (NED) is operated by the Jet Propulsion Laboratory, California Institute of Technology, under contract with the National Aeronautics and Space Administration. 
This work was supported by NASA under FERMI grant GSFC \#21078/FERMI08-0051 and the NRAO under Student Observing Support Award GSSP10-011.  Additional support provided by the Naval Research Laboratory.
\end{acknowledgments}

\bigskip 

\end{document}